\documentclass[letterpaper,11pt]{scrartcl}

\usepackage{paper}

\title{Protein structure prediction guided by cross-linking restraints --- A systematic evaluation
  of the impact of the cross-linking spacer length}
\author[1,*]{Tommy Hofmann}
\author[2,*]{Axel W. Fischer}
\author[2]{Jens Meiler}
\author[1,3]{Stefan Kalkhof}
\affil[1]{Department of Proteomics, Helmholtz-Centre for Environmental Research --- UFZ, Leipzig, Germany}
\affil[2]{Department of Chemistry and Center for Structural Biology, Vanderbilt University,
  Nashville, USA}
\affil[3]{Department of Bioanalytics, University of Applied Sciences and Arts of Coburg, Coburg, Germany}
\affil[*]{Contributed equally to this article}

\date{}

\begin{document}

\maketitle

\begin{abstract}

  Recent development of high-resolution \gls{ms} instruments enables chemical \gls{xl} to become a
  high-throughput method for obtaining structural information about proteins. Restraints derived
  from \gls{xl}-\gls{ms} experiments have been used successfully for structure refinement and
  protein-protein docking. However, one formidable question is under which circumstances
  \gls{xl}-\gls{ms} data might be sufficient to determine a protein's tertiary structure \emph{de
    novo}? Answering this question will not only include understanding the impact of
  \gls{xl}-\gls{ms} data on sampling and scoring within a \emph{de novo} protein structure
  prediction algorithm, it must also determine an optimal cross-linker type and length for protein
  structure determination. Whereas a longer cross-linker will yield more restraints, the value of
  each restraint for protein structure prediction decreases as the restraint is consistent with a
  larger conformational space.

  In this study, the number of cross-links and their discriminative power was systematically
  analyzed \emph{in silico} on a set of \num{2055} non-redundant protein folds considering Lys-Lys,
  Lys-Asp, Lys-Glu, Cys-Cys, and Arg-Arg reactive cross-linkers between \SI{1}{\angstrom} and
  \SI{60}{\angstrom}. Depending on the protein size a heuristic was developed that determines the
  optimal cross-linker length. Next, simulated restraints of variable length were used to \emph{de
    novo} predict the tertiary structure of fifteen proteins using the BCL::Fold algorithm. The
  results demonstrate that a distinct cross-linker length exists for which information content for
  \emph{de novo} protein structure prediction is maximized. The sampling accuracy improves on
  average by \SI{1.0}{\angstrom} and up to \SI{2.2}{\angstrom} in the most prominent
  example. \Gls{xl}-\gls{ms} restraints enable consistently an improved selection of native-like
  models with an average enrichment of \num{2.1}.

\end{abstract}

\section{Introduction}

``Structural Genomics'' --- the determination of the structure of all human proteins --- would have
profound impact on biochemical and biomedical research with direct implication to functional
annotation, interpretation of mutations, development of small molecule binders, enzyme design, or
prediction of protein-protein interaction \autocite{Baker2001}. Although significant progress
towards this goal has been made through X-ray crystallography and \gls{nmr} spectroscopy, tertiary
structure determination continues to be a challenge for many important human proteins. At present,
high-resolution structures exist for about \SI{5}{\percent} of all human proteins in the \gls{pdb}
\autocite{Berman2000}. For many uncharacterized human proteins, construction of a comparative model
is possible starting from the experimental structure of a related protein. Nevertheless, for about
\SI{60}{\percent} ($\sim7800$) of known protein families in the Pfam database \autocite{Punta2011}
not a single structure is deposited \autocite{Khafizov2014}. Many of these proteins will continue
to evade high-resolution protein structure determination.

Accordingly, researchers strive to develop alternative approaches. The most extreme approach
includes computational methods that predict the tertiary structure of proteins from their sequence
alone. Although computational methods are sometimes successful at the predicting the tertiary
structure of small proteins with up to one hundred residues \autocite{Moult2005}, for larger
proteins the size of the conformational space to be searched as well as the discrimination of
incorrectly folded models hinder structure prediction
\autocite{Bonneau2001,Bonneau2002,Bonneau2002a}.

However, recent studies demonstrate that combining \emph{de novo} protein structure prediction with
limited experimental data \autocite{Bowers2000, Alexander2008, Hirst2011, Fischer2015, Lindert2009,
  Lindert2012, Weiner2014}, \emph{i.e.}~experimental data that alone is insufficient to
unambiguously determine the fold of the protein, can yield accurate models for larger proteins. The
structural restraints in those studies were acquired using \gls{epr} spectroscopy
\autocite{Alexander2008, Hirst2011, Fischer2015}, \gls{em} \autocite{Lindert2009, Lindert2012}, or
\gls{nmr} spectroscopy \autocite{Weiner2014}.

As an alternative technique, \gls{xl} in combination with \gls{ms} can be applied to obtain
distance restraints, which can be used to guide protein structure prediction
\autocite{Petrotchenko2010, Sinz2006, Rappsilber2011, Young2000}. Using bifunctional reagents with
a defined length, functional groups within the protein can be covalently bridged in a native-like
environment. Thus, it is possible to determine an upper limit for the distance between those
residues after enzymatic proteolysis and identification of cross-linked peptides.

This method allows for a fast analysis of protein structures in a native-like environment at a low
concentration and can even be applied to high molecular weight proteins \autocite{Lasker2012},
membrane proteins \autocite{Jacobsen2006}, or highly flexible proteins \autocite{Kalkhof2005}. If
combined with affinity purification it becomes possible to study proteins inside the cell
\autocite{Sinz2010}. Currently, the \gls{xl}-\gls{ms} technology is rapidly gaining importance
driven by the \gls{lc}-\gls{ms} instrument development, the generation of advanced analysis software
\autocite{Tinnefeld2014}, and the direct integration in protein structure prediction workflows
\autocite{Kahraman2013, Kalkhof2010, Leitner2010}. Furthermore, hundreds of different cross-linking
reagents with different spacer lengths, reactivities, and features for specific enrichment and
improved detectability are now commercially available \autocite{Zybailov2013}.

However, whereas the potential to combine \gls{xl}-\gls{ms} and computational modeling has been
frequently demonstrated and many technical problems of \gls{xl}-\gls{ms} have been solved, several
central questions have not yet been evaluated systematically.

\begin{enumerate}[(i)]
\item Cross-linking reagents are available with a spacer length ranging from \SI{0}{\angstrom} to
  more than \SI{35}{\angstrom}. Whereas longer reagents are likely to provide more distance
  restraints, shorter cross-links have higher information content in \emph{de novo} structure
  prediction as the conformational search space is more restricted. Thus, the question arises,
  which cross-linker spacer length supports structure prediction best?
\item Cross-linking results are often used to confirm already existing structures. However, what is
  the average gain in model accuracy and selection of correct models when using cross-linking data
  in conjunction with \emph{de novo} protein structure prediction?
\item Cross-linking reagents vary in reactivity towards different functional groups present in
  different amino acids. For \emph{de novo} protein structure prediction, what is the gain of using
  additionally cross-linkers with different reactivities?
\end{enumerate}

In this study, we simulated cross-linking experiments on more than \num{2000} non-redundant protein
structures to determine the number of possible and structurally relevant cross-links depending on
the size of the protein as well as on the length and reactivity of the applied cross-linking
reagents. We then tested the impact of cross-linking restraints on \emph{de novo} protein structure
prediction for fifteen selected proteins.

\section{Materials and methods} \label{sec:xlink_methods}

\subsection{Software and databases}

A subset of the \gls{pdb} containing \num{2055} non-redundant protein structures was downloaded
from the PISCES server (Version 08.2012) \autocite{Wang2003}. This \gls{pdb} subset was created by
filtering all available structures with a resolution of at least \SI{1.6}{\angstrom}, a maximum
sequence identity of \SI{20}{\percent}, and an R-factor cutoff of \num{0.25}. Euclidean distances
and shortest \gls{sas} path lengths between $\mathrm{C_\beta}$-$\mathrm{C_\beta}$, Lys-Nz-Lys-Nz,
Lys-Nz-Asp-$\mathrm{C_\gamma}$, and Lys-Nz-Glu-$\mathrm{C_\delta}$, as well as
Arg-$\mathrm{N_{H2}}$-Arg-$\mathrm{N_{H2}}$ and Cys-$\mathrm{S_G}$-Cys-$\mathrm{S_G}$ atom pairs
with a maximum intramolecular distance of \SI{60}{\angstrom} were determined through the command
line version of Xwalk \autocite{Kahraman2011}.

\subsection{Generation of sequence dependent distance
  functions} \label{sec:xlink_distance_function}

Tables containing the Euclidean distances and the sequence separation between cross-linking target
amino acids
\begin{inparaenum}[(i)]
\item Lys-Lys,
\item Lys-Asp,
\item Lys-Glu,
\item Arg-Arg, and
\item Cys-Cys
\end{inparaenum}
were generated. Amino acid pair distances were sorted into \SI{2.5}{\angstrom} bins. The total
number of observed pairs for each sequence and Euclidean distance was counted. Based on the result
an approximation of the distance distribution for every sequence distance was created. The median
of the distribution was determined. A logarithmic function was calculated as a regression curve in
the form $E_{\mathit{med}} = a \times ln(S) + b$ to correlate the sequence separation $S$ to the
median Euclidean distances $E_{\mathit{med}}$.

\begin{figure}
  \centering
  \includegraphics[width=0.8\textwidth]{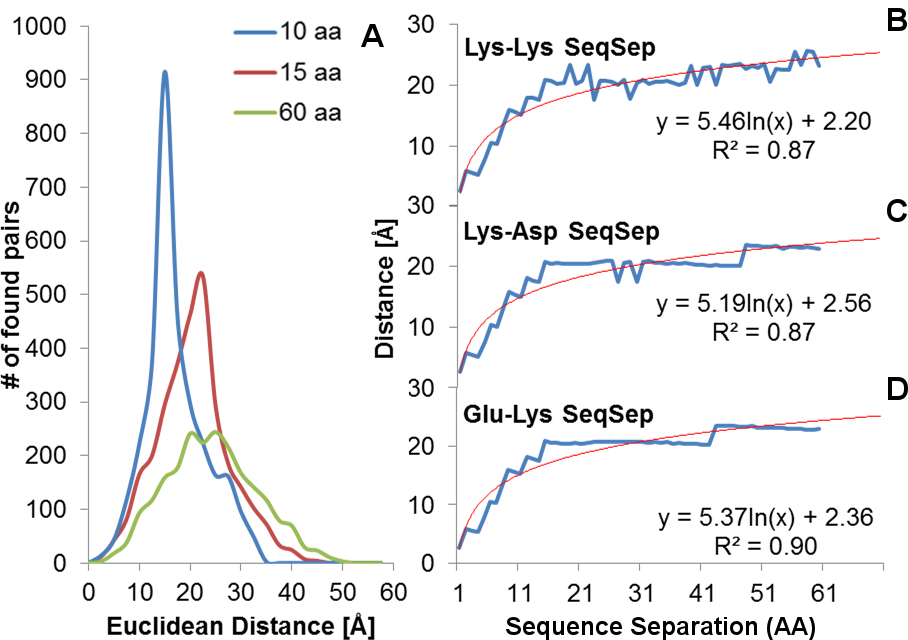}
  \caption[Residue pair distance distributions]{\textbf{Residue pair distance distributions.} (A)
    Distribution of the number of Lys-Lys pairs in respect to their Euclidean distance and (B-D)
    functions representing the relationship between sequence and spatial distance approximated by
    method of least squares to a logarithmic equation for (B) Lys-Lys, (C) Lys-Glu, and (D)
    Lys-Asp.}
  \label{fig:xlink_res_pair_distributions}
\end{figure}

\subsection{Calculation of the amino acid side-chain length}

Based on the structure of calmodulin (\gls{pdb} entry 2KSZ) the average $\mathrm{C_\beta}$-Nz,
$\mathrm{C_\beta}$-$\mathrm{C_\gamma}$, $\mathrm{C_\beta}$-$\mathrm{C_\delta}$,
$\mathrm{C_\beta}$-$\mathrm{N_{H2}}$, and $\mathrm{C_{\beta}}$-$\mathrm{S_G}$ distances of the
side-chains of lysine, aspartic acid, glutamic acid, arginine, and cysteine were determined to be
\SI{4.5}{\angstrom}, \SI{2.3}{\angstrom}, \SI{3.6}{\angstrom}, \SI{5.1}{\angstrom}, and
\SI{1.8}{\angstrom}, respectively.

\subsection{Distinguishing impossible, possible and structurally valuable cross-links}

Cross-linker spacer lengths between \SI{1}{\angstrom} and \SI{60}{\angstrom} distances were
evaluated and classified in either \begin{inparaenum}[(i)] \item impossible cross-links, meaning
  that the distance between the $\mathrm{C_\beta}$-atoms of the cross-linked amino acids exceeds
  the sum of the spacer lengths and the side-chain lengths, or \item possible cross-links, meaning
  that the $\mathrm{C_\beta}$-$\mathrm{C_\beta}$ distance is below the sum of the spacer lengths
  and side-chain lengths. \end{inparaenum} The latter group was subdivided into cross-links
potentially useful for structure determination (valuable cross-links) and those that are unlikely
to contribute much information (non-valuable cross-links). We defined cross-links as valuable if
the spacer length is shorter than the median distance expected for the given sequence separation by
the equations derived in \fref{sec:xlink_distance_function} (see also
\Fref{fig:xlink_res_pair_distributions}). For these calculations, all proteins were grouped into
\SI{2.5}{\kilo \dalton} bins. The calculations were performed for cross-linker lengths from
\SIrange{1}{60}{\angstrom} with a step size of \SI{1}{\angstrom}.

\subsection{Estimation of optimal spacer lengths for a given protein molecular weight}

\begin{wrapfigure}{R}{0.48\textwidth}
  \centering
  \includegraphics[width=0.46\textwidth]{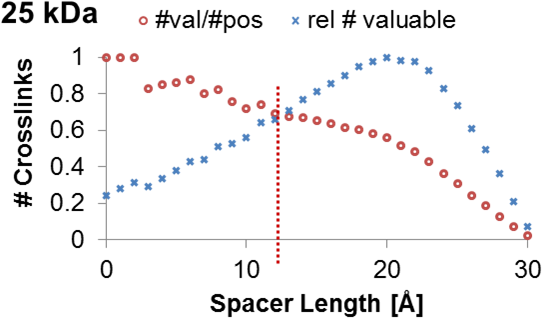}
  \caption[Cross-link yield in dependence of the spacer length]{\textbf{Cross-link yield in
      dependence of the spacer length.} Behavior of valuable and possible cross-links in the
    \gls{mw} bin \SI{25}{\kilo \dalton} and localization of the optimal spacer length. Shown is the
    number of valuable cross-links for every tested spacer length in red. These values are
    normalized to a dimension spanning \num{1}. Blue points show the share of valuable cross-links
    among the physical possible ones. The dotted line meets the intersection of both curves and
    represents the optimal spacer length where the best ratio between valuable and possible
    cross-links is attained and the number of valuable cross-links is maximized in respect to this
    ratio.}
  \label{fig:xlink_yield_length}
\end{wrapfigure}

Over all proteins in each \gls{mw} bin, the total number of possible distance pairs
($\#\text{possible}$) as well as the number of distance pairs useful for structure determination
($\#\text{valuable}$) were computed for each cross-linker spacer length. Furthermore, the maximum
number of valuable cross-links observed for all spacer lengths
($\#\text{valuable}_{\mathit{max}}$) was determined. For each \gls{mw} bin the ratios
($\frac{\#\text{valuable}}{\#\text{possible}}$) and
($\frac{\#\text{valuable}}{\#\text{valuable}_{\mathit{max}}}$) were plotted as a function of
the cross-linker spacer length. The optimal cross-linker length for each \gls{mw} bin was
approximated as intersection points of the two functions using a local regression
(\Fref{fig:xlink_yield_length}). The estimated values for the optimal cross-linker spacer length
were plotted as a function of the \gls{mw} and were fitted using a cubic regression curve. The
script used for the calculation is available at \url{http://www.ufz.de/index.php?en=19910}.

\subsection{Simulation of cross-linking restraints}

Seventeen proteins with known tertiary structure determined via X-ray crystallography (resolution
of \textless \SI{1.9}{\angstrom}) were selected from the data set of structures as test cases to
evaluate the influence of cross-linking restraints on \emph{de novo} protein structure
prediction. To thoroughly benchmark the algorithm, the benchmark set covers a wide range of protein
topologies and structural features. The sequence lengths of the proteins range from
\numrange{105}{303} residues, the number of \glspl{sse} ranges from \numrange{5}{19} with varying
$\mathrm{\alpha}$-helical and $\mathrm{\beta}$-strand content (\Fref{tab:xlink_benchmark_set}). For
these proteins, all solvent accessible surface $\mathrm{C_\beta}$-$\mathrm{C_\beta}$ distances
between target amino acids in the structure which were within the range of either homobifunctional
Lys-reactive cross-linkers or heterobifunctional Lys-Asp/Glu reactive cross-linkers were determined
through Xwalk. For the predicted optimal cross-linker length (read above) and spacer lengths of
\SIlist{2.5;7.5;17.5;30.0}{\angstrom} lists of structurally possible cross-links were generated.

\begin{table}
  \small
  \centering
  \begin{tabular}{llccrrrr}
  \toprule
  Protein & Uniprot & resolution & weight & length & Lysine  & $\mathrm{\alpha}$-helix & $\mathrm{\beta}$-strand \\
  \midrule
  1HRC      & P00004  & \SI{1.9}{\angstrom}                & \SI{12368}{\dalton}                     & 105                      & \SI{18}{\percent}                   & \SI{40}{\percent}               & \SI{1}{\percent}                \\
  3IV4      & Q7A6S3  & \SI{1.5}{\angstrom}                & \SI{13235}{\dalton}                     & 112                      & \SI{6}{\percent}                    & \SI{49}{\percent}               & \SI{25}{\percent}               \\
  1BGF      & P42228  & \SI{1.5}{\angstrom}                & \SI{14504}{\dalton}                     & 124                      & \SI{5}{\percent}                    & \SI{79}{\percent}               & \SI{1}{\percent}                \\
  1T3Y      & Q14019  & \SI{1.2}{\angstrom}                & \SI{15835}{\dalton}                     & 141                      & \SI{9}{\percent}                    & \SI{35}{\percent}               & \SI{29}{\percent}               \\
  3M1X      & C4LXT9  & \SI{1.2}{\angstrom}                & \SI{15882}{\dalton}                     & 138                      & \SI{7}{\percent}                    & \SI{25}{\percent}               & \SI{28}{\percent}               \\
  1X91      & Q9LNF2  & \SI{1.5}{\angstrom}                & \SI{16419}{\dalton}                     & 153                      & \SI{7}{\percent}                    & \SI{76}{\percent}               & \SI{0}{\percent}                \\
  1JL1      & P0A7Y4  & \SI{1.3}{\angstrom}                & \SI{17483}{\dalton}                     & 155                      & \SI{7}{\percent}                    & \SI{34}{\percent}               & \SI{30}{\percent}               \\
  1MBO      & P02185  & \SI{1.6}{\angstrom}                & \SI{17980}{\dalton}                     & 153                      & \SI{12}{\percent}                   & \SI{77}{\percent}               & \SI{0}{\percent}                \\
  2QNL      & Q11XA0  & \SI{1.5}{\angstrom}                & \SI{19218}{\dalton}                     & 162                      & \SI{5}{\percent}                    & \SI{70}{\percent}               & \SI{2}{\percent}                \\
  2AP3      & Q8NX77  & \SI{1.6}{\angstrom}                & \SI{23190}{\dalton}                     & 199                      & \SI{23}{\percent}                   & \SI{81}{\percent}               & \SI{0}{\percent}                \\
  1J77      & Q9RGD9  & \SI{1.5}{\angstrom}                & \SI{24226}{\dalton}                     & 209                      & \SI{8}{\percent}                    & \SI{62}{\percent}               & \SI{1}{\percent}                \\
  1ES9      & Q29460  & \SI{1.3}{\angstrom}                & \SI{25876}{\dalton}                     & 232                      & \SI{3}{\percent}                    & \SI{41}{\percent}               & \SI{11}{\percent}               \\
  3B5O      & D0VWS1  & \SI{1.4}{\angstrom}                & \SI{27506}{\dalton}                     & 244                      & \SI{3}{\percent}                    & \SI{71}{\percent}               & \SI{0}{\percent}                \\
  1QX0      & P0A2Y6  & \SI{2.3}{\angstrom}                & \SI{32821}{\dalton}                     & 293                      & \SI{7}{\percent}                    & \SI{38}{\percent}               & \SI{20}{\percent}               \\
  2IXM      & Q15257  & \SI{1.5}{\angstrom}                & \SI{34798}{\dalton}                     & 303                      & \SI{7}{\percent}                    & \SI{60}{\percent}               & \SI{3}{\percent}                \\
  FGF2      & P09038  & \SI{1.5}{\angstrom}                & \SI{17859}{\dalton}                     & 145                      & \SI{10}{\percent}                   & \SI{9}{\percent}                & \SI{34}{\percent}               \\
  P11       & P60903  & \SI{2.0}{\angstrom}                & \SI{11071}{\dalton}                     & 95                       & \SI{13}{\percent}                   & \SI{63}{\percent}               & \SI{3}{\percent}                \\
  \bottomrule
\end{tabular}

  \caption[Proteins used for the cross-link spacer length benchmark]{\textbf{Proteins used for the
      cross-link spacer length benchmark.} The fifteen proteins for the benchmark set were selected
    from high-resolution structures deposited in the \gls{pdb} with varying content of lysines. The
    structures were selected to cover a wide range of the structural features sequence length as
    well as percentage of residues within $\mathrm{\alpha}$-helices and $\mathrm{\beta}$-strands.}
  \label{tab:xlink_benchmark_set}
\end{table}

For the two proteins horse heart cytochrome c (\gls{pdb} entry 1HRC) and oxymyoglobin (\gls{pdb}
entry 1MBO) restraints were also derived from published cross-linking \gls{ms} experiments
deposited in the \gls{xl} database \autocite{Kahraman2013}. Experimental cross-linking data of FGF2
(\gls{pdb} entry 1FGA) and P11 (\gls{pdb} entry 4HRE) were derived from Young \emph{et
  al.}~\autocite{Young2000} and Schulz \emph{et al.}~\autocite{Schulz2007}, respectively.

\subsection{Translating cross-linking data into structural restraints}

\begin{figure}[h]
  \includegraphics[width=\textwidth]{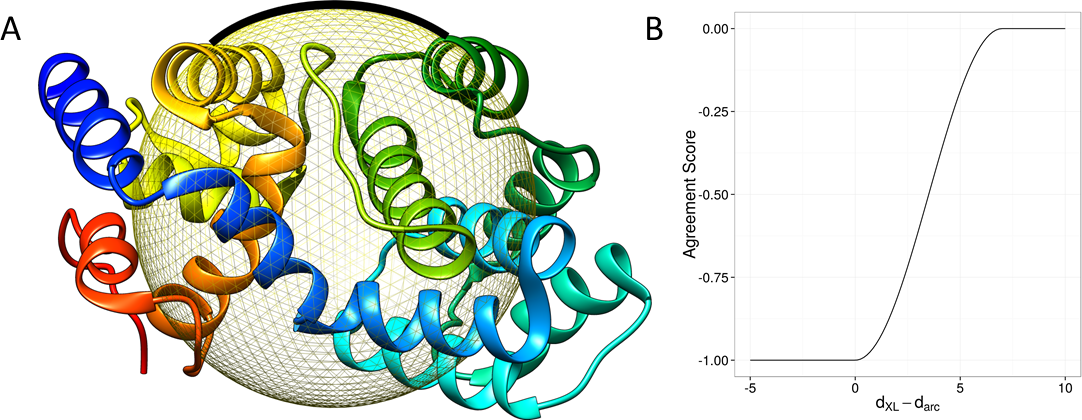}
  \caption[Implicit translation from cross-linking data into structural
  restraints]{\textbf{Implicit translation from cross-linking data into structural restraints.}
    Explicit simulation of the cross-linker conformation is computationally expensive and
    prohibitive for use in a rapid scoring function required for protein structure
    prediction. Instead, the cross-linker conformation and the path crossed by the cross-linker
    were approximated through computing the arc length connecting the two cross-linked residues
    (A). The agreement of a model with cross-linking data was evaluated by computing the difference
    between the arc length ($d_{\mathit{arc}}$) and the cross-linker length
    ($d_{\mathit{xl}}$). The agreement of the model with the cross-linking data is quantified with
    a score between \num{-1} and \num{0}, with \num{-1} being the best agreement and \num{0} being
    the worst agreement (B).}
  \label{fig:xlink_translation}
\end{figure}

Explicitly rebuilding coordinates for a cross-link is comparable to solving the loop closure
problem \autocite{Canutescu2003}. During \emph{de novo}, protein structure prediction the
cross-link would have to be reconstructed each time the conformation of the protein changes. In a
typical \gls{mc} simulation with a maximum of \num{12000} \gls{mc} steps per model and \num{5000}
models for each protein this would result in a maximum number of \num{60} million attempts to build
the cross-link, which is too resource demanding for usage in \emph{de novo} protein structure
prediction. Therefore, we developed a fast approach to estimate the chance that a particular model
fulfills a \gls{xl}-\gls{ms} restraint. The surface path of a cross-link is approximated by laying
a sphere around the protein structure and computing the arc length between the cross-linked
residues (\Fref{fig:xlink_translation}). The geometrical center of the protein structure is used as
the center of the sphere. If takeoff and landing point have different distances to the center of
the sphere, the longer distance is used as the radius. During the protein structure prediction
process, the side-chains of the residues are not modeled explicitly but represented on a simplified
way through a 'super atom'. While this simplification vastly reduces the computational demand of
the algorithm, it also adds additional uncertainty due to the unknown side-chain conformations. The
agreement of the model with the cross-linking data is quantified by comparing the distance between
the cross-linker lengths ($l_{\mathit{XS}} + l_{\mathit{SS1}} + l_{\mathit{SS2}}$) with the
computed arc lengths ($d_{arc}$), with \num{-1} being the best agreement and \num{0} being the
worst agreement. To account for the uncertainty of side-chain conformations a cosine-transition
region of \SI{7}{\angstrom} was introduced (\Fref{fig:xlink_translation}).

\subsection{Structure prediction protocol for the benchmark set}

\begin{wrapfigure}{R}{0.5\textwidth}
  \centering
  \includegraphics[width=0.48\textwidth]{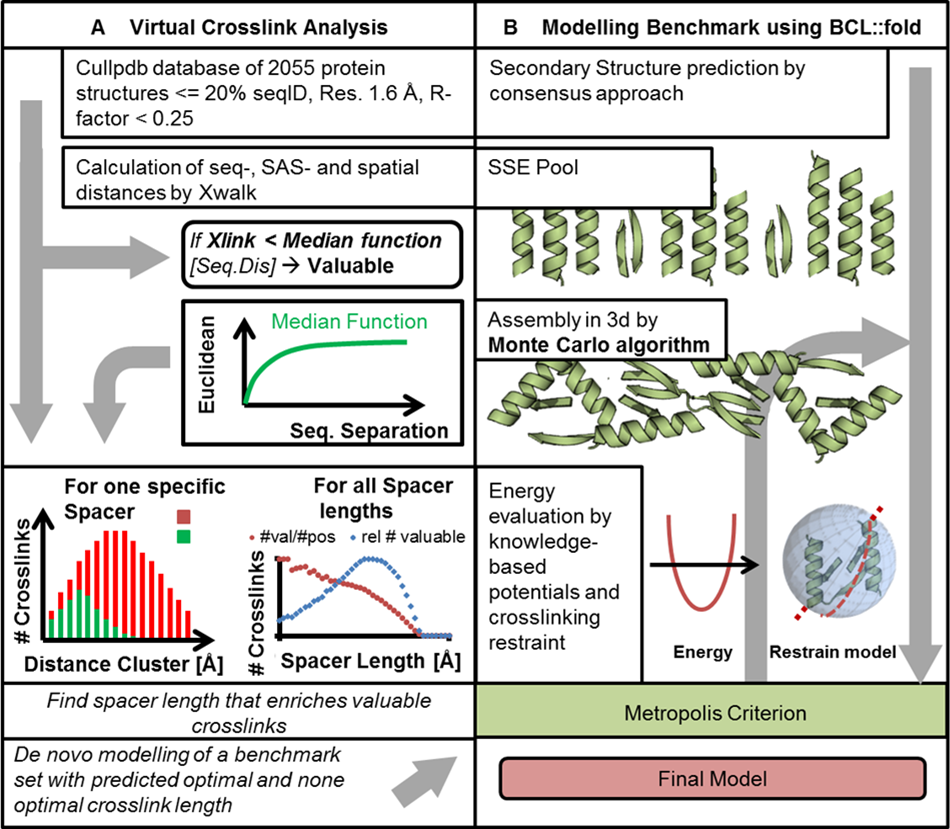}
  \caption[Spacer length and protein structure prediction workflow]{\textbf{Spacer length and
      protein structure prediction workflow.} Workflow for (A) the prediction of optimal
    cross-linker spacer length and (B) for \emph{de novo} protein structure prediction using
    BCL::Fold. (A) Workflow for the prediction of the optimal spacer length depending on the
    \gls{mw} of the protein of interest. (B) Workflow for \emph{de novo} protein structure
    prediction using BCL::Fold. \Glspl{sse} are predicted using the \gls{sse} prediction methods
    PsiPred and Jufo9D. A \gls{mcm} algorithm subsequently searches the conformational space for
    the structure with most favorable score.}
  \label{fig:xlink_workflow}
\end{wrapfigure}

The protein structure prediction protocol is based on the BCL::Fold protocol for soluble proteins
\autocite{Karakas2012}. In a preparatory step, the glspl{sse} are predicted using the \gls{sse}
prediction methods PsiPred \autocite{Jones1999} and Jufo9D \autocite{Leman2013} and an \gls{sse}
pool is created. Subsequently a \gls{mcm} energy minimization algorithm draws random glspl{sse}
from the predicted \gls{sse} pool and places them in the three-dimensional space
(\Fref{fig:xlink_workflow}). Random transformations like translation, rotation or shuffling of
glspl{sse} are applied. After each \gls{mc} step the energy of the resulting model is evaluated
using knowledge-based potentials which, among others, evaluate the packing of glspl{sse}, exposure
of residues, radius of gyration, pairwise amino acid interactions, loop closure geometry and amino
acid clashes \autocite{Wotzel2012}. Based on the energy difference to the previous step and the
simulated temperature a Metropolis criterion decides whether to accept or reject the most recent
change.

The protein structure prediction protocol is broken into multiple stages, which differ regarding
the granularity of the transformations applied, and the emphasis of different scoring terms. The
first five stages apply large structural perturbations, which can alter the topology of the
protein. Each of the five stages lasts for a maximum of \num{2000} \gls{mc} steps. If an
energetically improved structure has not been generated within the previous \num{400} \gls{mc}
steps, the stage terminates. Over the course of the five assembly stages, the weight of clashing
penalties in the total score is ramped up as \numlist{0;125;250;375;500}.

The five protein assembly stages are followed by a stage of structural refinement. This stage lasts
for a maximum number of \num{2000} \gls{mc} steps and terminates if no energetically improved model
is sampled for \num{400} \gls{mc} steps in a row. Unlike the assembly stages, the refinement stage
only consists of small structural perturbations, which will not drastically alter the topology of
the protein model.

Through multiple prediction runs with different score weights, the optimal contribution of the
cross-linking score to the total score was determined to be
\SIrange{40}{50}{\percent}. Consequently, the weight for the scoring term evaluating the agreement
of the model with the cross-linking data was set to \num{300} over all six stages, which ensures
that the cross-linking score contributes between \SIlist{40;50}{\percent} to the total score.

\subsection{\emph{De novo} folding simulations without and with cross-linking restraints}

To evaluate the influence of cross-linking restraints on protein structure prediction accuracy,
each protein was folded in the absence and in the presence of Lys-Lys, Lys-Glu, and Lys-Asp
cross-linking restraints. Independent structure prediction experiments were performed for the
predicted optimal as well as two shorter and two longer cross-linker spacer lengths each of the
five spacer lengths (\Fref{tab:xlink_num_lys_lys}). Additionally, predictions were performed using
combination of all spacer lengths as well as using restraints obtained by the optimal spacer length
of all three cross-linker reactivities. For the two proteins of which experimentally determined
cross-linking data were available, protein structure prediction was additionally performed for the
experimentally determined restraints. For each protein and cross-linker length used, \num{5000}
models were sampled in independent \gls{mcm} trajectories. Due to the randomness of the employed
\gls{mc} algorithm, ten sets of \num{5000} models were sampled for each protein without
restraints. Improvements in prediction accuracy can be compared to the standard deviations to
identify statistically significant improvements (\Fref{tab:xlink_prediction_results}).

\subsection{Metrics for comparing calculating model accuracy and enrichment}

The prediction results were evaluated using the \gls{rmsd100} \autocite{Carugo2001} and enrichment
\autocite{Wotzel2012, Fischer2015} metrics. The \gls{rmsd100} metric was used to quantify the
sampling accuracy by computing the normalized root-mean-square-deviation between the backbone atoms
of the superimposed model and native structure. The enrichment metric was used to quantify the
discrimination power of the scoring function by computing which percentage of the most accurate
models can be selected by the scoring function. The enrichment metric is used to assess the
influence of the cross-linking restraints to discriminate among the sampled models. First, the
models of a given set $S$ are sorted by their \gls{rmsd100} relative to the native structure. The
\SI{10}{\percent} of the models in $S$ with the lowest \gls{rmsd100} are assigned to subset $P$
(positives) and the remaining \SI{90}{\percent} of the models are assigned to subset $N$
(negatives). Second, the models in $S$ are sorted by their BCL score. The \SI{10}{\percent} of the
models in $S$ with the best score are assigned to subset $\mathit{FS}$ (favorable score). The
intersection $\mathit{TP} = \mathit{FS} \cap P$ contains the most accurate models which the scoring
function can select (true positives). The enrichment
$e = \frac{\#\mathit{TP}}{\#P} \times \frac{\#P + \#N}{\#P}$ of the most accurate models the
scoring function can select. In order to reduce the influence of the sampling accuracy on the
enrichment values, the positive models are considered the \SI{10}{\percent} of the models with the
lowest \gls{rmsd100} and $\frac{\#P + \#N}{\#P}$ is fixed at a value of \num{10.0}. Therefore, the
enrichment ranges from \numrange{0.0}{10.0}, with a score of \num{1.0} indicating random selection
and a value above \num{1.0} indicating that the scoring function enriches for native-like models.

\section{Results}

\subsection{Creation of an \emph{in silico} cross-linking database}

We performed \emph{in silico} cross-linking experiments on \num{2055} non-redundant
proteins. Covering an \gls{mw} range from \SIrange{1.4}{139}{\kilo \dalton}, \SI{59}{\percent} of
the proteins have an \gls{mw} below \SI{25}{\kilo \dalton}. For each of those proteins all Lys-Lys,
Lys-Asp, and Lys-Glu sequence and Euclidean distances as well as the \gls{sas} distance between the
$\mathrm{C_\beta}$-atoms were determined. Thus, the resulting database contained information on
\num{391902} Lys-Lys, \num{395815} Lys-Glu, and \num{360101} Lys-Asp pairs which built the basis
for the determination of the number of possible cross-links, cross-links useful for structure
prediction, and finally for the prediction of the optimal cross-linker length for studying a
selected protein (\Fref{fig:xlink_workflow}).

\subsection{Estimation of the possible cross-links per protein}

Next we estimated how many and which of the distances could be cross-linked with a cross-linker of
a given length and specificity. We considered cross-links possible if the sum of the spacer length
and the length of the two connected side-chains ($\mathrm{C_\beta}$-$\mathrm{C_\beta}$,
Lys-Nz-Lys-Nz, Lys-Nz-Asp-$\mathrm{C_\gamma}$, or Lys-Nz-Glu-$\mathrm{C_\delta}$) is longer than
the $\mathrm{C_\beta}$-$\mathrm{C_\beta}$-\gls{sas}-distance between the amino acids. As the
lengths of the side-chains of Lys ($\mathrm{C_\beta}$-Nz), Asp ($\mathrm{C_\beta}$-Oz), and Glu
($\mathrm{C_\beta}$-Oz) \SI{4.5}{\angstrom}, \SI{2.4}{\angstrom}, and \SI{3.6}{\angstrom} were
used, which were determined as average values from the crystal structure of calmodulin (\gls{pdb}
entry 1CLL). \emph{In silico} cross-linking experiments were conducted for all of the \num{2055}
proteins using homobifunctional Lys-Lys-reactive, as well as heterobifunctional (Lys-Asp- and
Lys-Glu-reactive) cross-linking reagents with lengths from \SIrange{1}{60}{\angstrom} (step size
\SI{1}{\angstrom}).

To draw conclusions from the correlation of this \emph{in silico} cross-linking experiments to the
\gls{mw} of the studied proteins the proteins were grouped into \num{45} bins with a step size of
\SI{2.5}{\kilo \dalton}. For example, a protein with a \gls{mw} in the range of
\SIrange{25}{27.5}{\kilo \dalton} contains on average \num{15.1} Lys, \num{14.4} Asp, and
\num{16.7} Glu. On average, \num{182} Lys-Lys, \num{173} Lys-Glu, and \num{144} Lys-Asp cross-links
exist per protein within this specific \gls{mw} bin. Theoretically, all of those could be
cross-linked with a cross-linker of \SI{60}{\angstrom}. In contrast by utilization of cross-linkers
of \SI{13}{\angstrom} (as e.g. BS3) only about \SI{33}{\percent} of the cross-links are formed
\emph{in silico}. When going to a cross-linker of length of \SI{1}{\angstrom} (e.g. close to EDC),
only \SI{10}{\percent} of all possible amino acid pairs are linked.

\subsection{Estimation of structurally relevant cross-links}

In protein structure prediction approaches, the enrichment of low RMSD structures among thousands
of generated models is crucial. Therefore, we hypothesized that restraints that are valuable for
structure prediction will reduce the conformational search space substantially. For the present
study, we classify a cross-linking restraint as useful for structure prediction if it discriminate
at least \SI{50}{\percent} of all possible conformations. Thus, in a second step each of the
possible cross-links was evaluated in terms of its potential to discriminate at least
\SI{50}{\percent} of incorrect structures (useful for structure determination) or whether the
cross-linked amino acids are so close in sequence that it can be derived from sequence separation
that the distance can be bridged by the cross-linker independently of the protein's structure (not
useful for structure determination).

In order to develop a stringent measure for usefulness we did not simply assume the maximum
distance that can be bridged by an amino acid chain of a certain length. Instead, the Euclidean
distance distributions for Lys-Lys, Lys-Glu, and Lys-Asp were computed for the sequence separations
ranging from \numrange{1}{60} amino acids within our database of protein structures. For example,
in the more than \num{2000} analyzed structures there are \num{3132} Lys-Lys pairs, which are
separated by ten amino acids. For this sequence distance Euclidean distances bins of
\SI{2.5}{\angstrom} were defined in which the occurrences of residue pairs were counted. The pairs
were present in bins ranging from \SIrange{2.5}{35.0}{\angstrom}. As the median distance, we found
\SI{15.5}{\angstrom}. For the same sequence distance the distribution of Lys-Glu (\num{3336} pairs)
and Lys-Asp (\num{3010} pairs) are quite similar and the median values were \SI{15.6}{\angstrom}
and \SI{15.3}{\angstrom}.

Similarly, for sequence separations of \num{15} amino acids we observed \num{3024} Lys-Lys pairs,
\num{3200} Lys-Glu pairs, and \num{2835} Lys-Asp pairs. The median values are \SI{20.8}{\angstrom},
\SI{20.9}{\angstrom}, and \SI{20.7}{\angstrom}, respectively. For sequence separations of \num{60}
amino acids, we observed \num{2167} Lys-Lys pairs, \num{2212} Lys-Glu pairs, and \num{2167} Lys-Asp
pairs. The median values are \SI{23.0}{\angstrom}, \SI{23.0}{\angstrom}, and \SI{23.0}{\angstrom},
respectively (\Fref{fig:xlink_res_pair_distributions}).

Approximating the proteins structures as spheres, we applied a logarithmic model to fit the
relationship between the sequence separation $S$ and the median Euclidean distance
$E_{\mathit{med}}$. We find
\begin{inparaenum}[(i)]
\item $E_{\mathit{Lys}-\mathit{Lys}} = 5.46 \times ln(S_{\mathit{Lys}-\mathit{Lys}}) + 2.2$,
\item $E_{\mathit{Lys}-\mathit{Glu}} = 5.37 \times ln(S_{\mathit{Lys}-\mathit{Glu}}) + 2.36$, and
\item $E_{\mathit{Lys}-\mathit{Asp}} = 5.19 \times ln(S_{\mathit{Lys}-\mathit{Asp}}) + 2.36$
\end{inparaenum}
for Lys-Lys, Lys-Glu, and Lys-Asp distances, respectively.

\begin{figure}[h]
  \centering
  \includegraphics{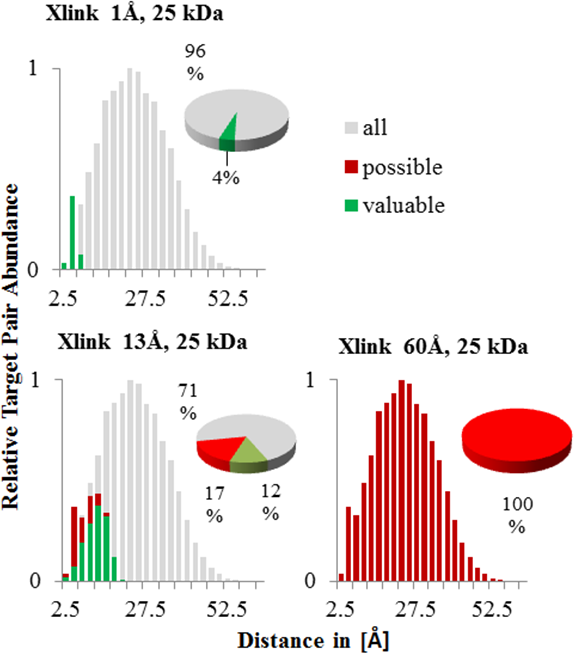}
  \caption[Lys-Lys pair distributions]{\textbf{Lys-Lys pair distributions.} Distribution of all
    possible and valuable Lys-Lys pairs for a \SIrange{25}{27.5}{\kilo \dalton} weight bin. Gray
    bars show all theoretical pairs in their specific distance cluster of $\pm$
    \SI{2.5}{\angstrom}. Red bars show pairs that could be connected in respect to their surface
    distance by a specific cross-link (here \SIlist{1;13;60}{\angstrom}) always including the
    side-chain contribution to the overall length. Green bars show pairs that are considered
    valuable by our proposed scoring function. Pie charts show the accumulated number of
    cross-links for every spacer length.}
  \label{fig:xlink_lys_lys_distribution}
\end{figure}

Secondly, using our derived functions constituting the $S/E$ relationships, we considered every
cross-link as of reasonable discriminative power, \emph{i.e.}~which fulfills the criterion that the
sum of the cross-linker spacer length and the average length of both contributing side-chains is
shorter than the median of the sequence/Euclidean-distance distribution. If we examine the
\SI{25}{\kilo \dalton} \gls{mw} bins of Lys-Lys targets with a \SI{1}{\angstrom} spacer cross-link
\num{1167} of the possible \num{22398} target pairs fulfilled this criterion and were considered as
of sufficient discriminative power (\Fref{fig:xlink_lys_lys_distribution}). These cross-links,
which represent \SI{4}{\percent} of all Lys-Lys distances we defined therefore as useful for
protein structure prediction. Application of a \SI{13}{\angstrom} spacer length results in
\num{2935} valuable target pairs (\SI{12}{\percent} of all Lys-Lys distances, see
\Fref{fig:xlink_lys_lys_distribution}). In contrast, a cross-linker with a spacer length of
\SI{60}{\angstrom} would allow to cross-link all distances. However, none of the cross-links would
have discriminative power for native-like models (\Fref{fig:xlink_lys_lys_distribution}). For the
proteins of the \SI{25}{\kilo \dalton} \gls{mw} bins the number of valuable cross-links as a
function of the cross-linker length has a log-normal character never exceeding a roughly
\SI{25}{\angstrom} spacer. The intermediate length of \SI{13}{\angstrom} resulted in an almost
equal contribution of valuable and structurally invaluable cross-linking pairs. Whereas
\SI{29}{\percent} of all possible reactive amino acid pairs are linked, \SI{12}{\percent} are
considered valuable according for structure prediction (\Fref{fig:xlink_lys_lys_distribution}).

\subsection{Prediction of molecular weight dependent optimal cross-linker spacer lengths}

Whereas usage of a short cross-linker will result in only a few but mostly structurally valuable
restraints, a longer cross-linker will yield more restraints but a lower ratio of valuable
restraints. Furthermore, the ratio of valuable restraints as well as the number of possible
restraints depends on the size of the protein. In agreement with prior studies regarding structural
modeling driven by sparse distance restraints \autocite{Havel1979}, we hypothesize that a
compromise between maximizing the portion of valuable cross-links compared to all cross-links which
can be formed with a given cross-linker length ($\frac{\#\text{valuable}}{\#\text{possible}}$)
and maximizing the relative number of achievable cross-links with any spacer length
($\frac{\#\text{valuable}}{\#\text{valuable}_{\mathit{max}}}$) might yield the best results.

Following our hypothesis, for each \gls{mw} bin we derived the optimal spacer length as the
intersection point of the two functions as it is shown exemplarily for \gls{mw} \SI{25}{\kilo
  \dalton} in \fref{fig:xlink_yield_length}.

\begin{wrapfigure}{R}{0.5\textwidth}
  \centering
  \includegraphics[width=0.48\textwidth]{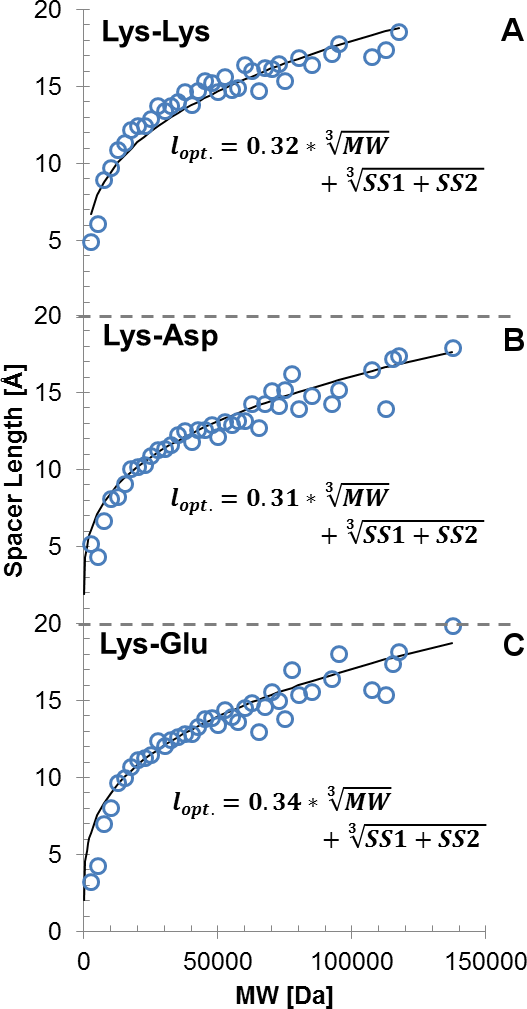}
  \caption[Relationship between sequence distance and Euclidean distance]{\textbf{Relationship
      between sequence distance and Euclidean distance.} Functions representing the relationship
    between sequence ($S$) and spatial distance ($E$). The equations approximated by method of
    least squares to a logarithmic equation for (A) Lys-Lys, (B) Lys-Glu, and (C) Lys-Asp.}
  \label{fig:xlink_reactivity_lengths}
\end{wrapfigure}

The derived optimal spacer lengths for Lys-Lys, Lys-Asp, and Lys-Glu were plotted as function of
the \gls{mw} (\Fref{fig:xlink_reactivity_lengths}). The relationship was fitted using a cube root
function. For our observable \gls{mw} sample space for Lys-Lys cross-links, all spacer lengths
reached dimensions between \SIlist{5.0;18.6}{\angstrom}. No optimal spacer length was further than
\SI{2.5}{\angstrom} separated from the regression curve. The average distance from the modeled
spacer lengths was \SI{0.7}{\angstrom}. The \gls{mw} term as well as the side-chain term has been
modeled as an exponential fraction in respect to the relation between volume and distances in
spherical objects.

Additionally, the optimal spacer lengths were also predicted for homobifunctional arginine and for
homobifunctional cysteine cross-linking reagents analogously to the procedure being described for
the homo- and heterobifunctional lysine-containing cross-links. Consistently, the optimal spacer
lengths depend on the \gls{mw} as well as the lengths of the cross-linked side-chains
$\mathit{SS1}$ and $\mathit{SS2}$ and could be calculated by
$l_{\mathit{opt}} [\si{\angstrom}] = k \times \sqrt[3]{\mathit{MW}} + \sqrt[3]{\mathit{SS1} +
  \mathit{SS2}}$. $k$ was determined to be \numlist{0.32;0.31;0.34;0.34;0.35} for Lys-Lys, Lys-Asp,
Lys-Glu, Arg-Arg, and Cys-Cys, respectively.

\subsection{Generation of \emph{in silico} and experimental cross-linking data for testing the
  effect of different spacer length for \emph{de novo} modeling}

To evaluate the effect of cross-linking data derived from experiments with different spacer length
we folded seventeen proteins \emph{de novo} with BCL::fold (\Fref{fig:xlink_workflow}). Thirteen
proteins were part of our data set while for four proteins experimental cross-linking data were
available (1MBO, 1HRC, 1FGA, and 4HRE) (\Fref{tab:xlink_benchmark_set}). All proteins have a
\gls{mw} in the range from \SIrange{13}{27}{\kilo \dalton}. Most structures were mainly
$\mathrm{\alpha}$-helical with fewer $\mathrm{\beta}$-strand \glspl{sse}. The
$\mathrm{\beta}$-strand content ranged from \SIrange{0}{51}{\percent}. The
$\mathrm{\alpha}$-helical content ranges from \SIrange{2}{81}{\percent}. The highest
$\mathrm{\beta}$-strand content showed 1LMI with also the fewest $\mathrm{\alpha}$-helices. The
portion of lysines was between \SIlist{3;23}{\percent}, which resulted in minimal \num{4} and
maximal \num{46} lysine residues per structure. For the two structures 1MBO and 1HRC, which were
studied experimentally, we used the published experimental data, which were obtained using DSG,
DSS/BS3, and DEST \autocite{Kahraman2013}. For 1MBO, there were \num{8} cross-links in total with
the \SI{11.4}{\angstrom} reagent BS3 four of them confirmed with the \SI{7.7}{\angstrom} reagent
DSG. For 1HRC, \num{48} cross-links were reported. \num{9} DSS, \num{31} BS3, \num{6} DSG, and
\num{9} with DEST (\SI{11}{\angstrom}). Six cross-links had been identified with different
cross-linking reagents. \num{18} BS3 cross-links were published for 1FGA \autocite{Young2000},
whereas \num{3} intramolecular BS3 cross-links were available for 1HRE \autocite{Schulz2007}. For
the thirteen proteins as well as for 1MBO and 1HRC, we predicted all cross-links, which are
possible with the predicted optimal cross-linker length as well as with two shorter and two longer
cross-linking reagents (\Fref{tab:xlink_num_lys_lys}) and used these data as restraints during
modeling (\Fref{fig:xlink_translation}).

\subsection{Cross-linking restraints improve the sampling accuracy of \emph{de novo} protein
  structure prediction}

\Gls{xl}-\gls{ms} data provides structural restraints that reduce the sampling space in \emph{de
  novo} structure determination. Thereby a fraction of incorrect conformations is excluded and the
sampling density in all other areas of the conformational space is increased. To evaluate the power
of cross-linking restraints to guide \emph{de novo} protein structure determination we computed the
\gls{rmsd100} \autocite{Carugo2001} values of the most accurate models for each protein for
structure prediction with and without cross-linking restraints. Using cross-linking restraints not
only increases the frequency with which accurate models are sampled, but the best models achieve an
accuracy not sampled in the absence of cross-linking data
(\Fref{tab:xlink_prediction_results}). Across all benchmark proteins, the accuracy of the best
models was, on average, \SI{6.6}{\angstrom} when no cross-linking data was used. By using
cross-linking, data for the spacer length deemed optimal the average \gls{rmsd100} value was
improved to \SI{5.6}{\angstrom}, which corresponds to two standard deviations. By using restraints
obtained for all five spacer lengths, the average accuracy of the best model improved to
\SI{5.2}{\angstrom}. For the proteins 1XQ0, 2IXM, and 3B50, even with cross-linking data, it was
not possible to sample a native-like conformation. We attribute this to limitations in the sampling
algorithm resulting in the native conformation not being part of the sampling space. For other
proteins, significant improvements could be observed. While the accuracy of the best models for
1ES9 and 1J77 was \SIlist{7.3;6.8}{\angstrom}, cross-linking restraints improved the accuracy to
\SIlist{5.7;4.5}{\angstrom}, respectively. For 1MBO, the accuracy could be improved from
\SI{7.1}{\angstrom} to \SI{4.2}{\angstrom} by using a combination of Lys-Glu/Asp reactive
cross-linkers (\Fref{fig:xlink_gallery}).

\subsection{Cross-linking restraints improve the discriminative power of the scoring function}

The ability of the scoring function to identify the most accurate models among the sampled ones was
quantified using the enrichment metric (see \fref{sec:xlink_methods}). Enrichments were computed
for proteins predicted without cross-linking data, for each spacer length and for all spacer
lengths combined. For protein structure prediction without cross-linking restraints an average
enrichment of \num{1.1} was measured, which is barely better than random selection. The scoring
function has almost no discriminative power. Using cross-linking restraints yielded by the optimal
spacer length improved the enrichment to \num{2.1} (\Fref{tab:xlink_prediction_results}), which
corresponds to three standard deviations. Using all five spacer lengths to obtain additional
restraints further improves the enrichment to \num{2.4}. The most significant improvement could be
observed for 1J77, for which the enrichment could be improved from \num{0.5} to \num{2.4}.

\subsection{The cross-linker length determines improvements in sampling accuracy and discrimination
  power}

The length of the cross-linker determines the number of obtainable restraints as well as their
information content \autocite{Alexander2008}. While a longer cross-linker is able to form more
cross-links and therefore yields a larger number of restraints, the longer cross-linker length can
be fulfilled by a larger number of conformations, reducing the discriminative power of the
restraint. In order to assess the influence of the cross-linker length, and therefore the number of
restraints and restraint distances, on the sampling accuracy and discrimination power, the protein
structure prediction protocol was conducted with restraints derived from different cross-linker
lengths.

\begin{figure}
  \centering
  \includegraphics[width=0.85\textwidth]{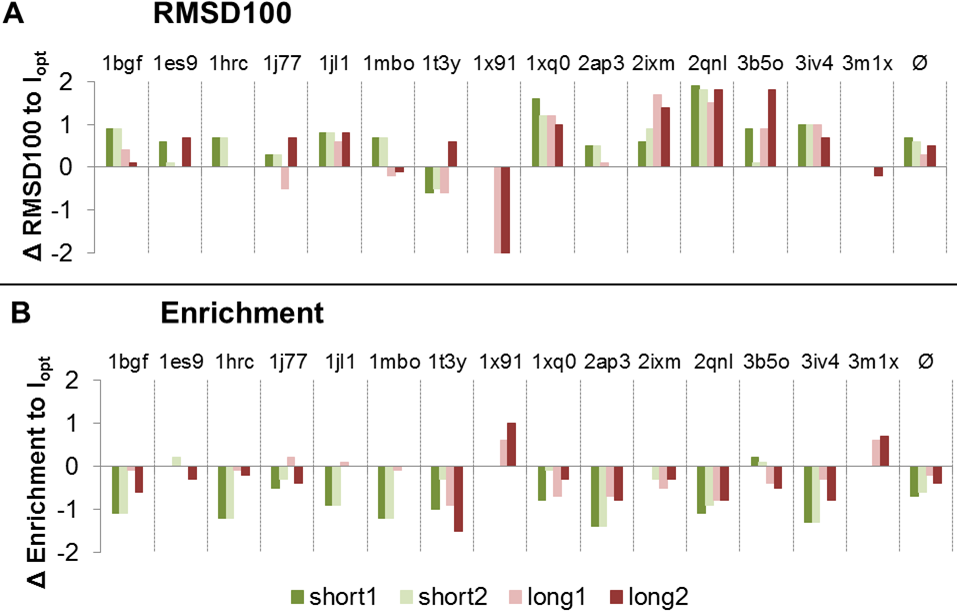}
  \caption[Protein structure prediction results for various spacer lengths]{\textbf{Protein
      structure prediction results for various spacer lengths.} Cross-linking data improve
    prediction accuracy and discrimination power. Using geometrical restraints derived from
    cross-linking experiments reduces the size of the conformational space, which needs to be
    searched for the conformation with the lowest free energy. This results in a higher likelihood
    of sampling accurate models and an improved discrimination power of the scoring function. Panel
    A compares the \gls{rmsd100} values of the most accurate model for structure prediction from
    different spacer lengths to the results for the optimal spacer length (horizontal line). Panel
    B compares the enrichments for different spacer lengths likewise.}
  \label{fig:xlink_results}
\end{figure}

The cross-linker lengths were separated into five groups: \emph{optimal}, which is the predicted
optimal cross-linker length, \emph{short1} and \emph{short2}, which are shorter cross-linker
lengths, and \emph{long1} and \emph{long2}, which are longer cross-linker lengths. The cross-linker
length predicted to be optimal yielded the most useful restraints for protein structure prediction
in terms of sampling accuracy and discriminative power. Across all proteins the average
\gls{rmsd100} values of the most accurate models for the optimal cross-linker length were
\SI{5.6}{\angstrom} --- an improvement by \SI{15}{\percent} --- while they were
\SIlist{6.3;6.2;5.9;6.1}{\angstrom} --- improvements by \SIlist{5;6;11;8}{\percent} --- for the
shorter and longer cross-linker lengths, respectively (\Fref{fig:xlink_results}). The longest
cross-linkers have a less significant impact on the sampling accuracy due to their ambiguity,
whereas the shortest cross-linkers failed to yield a sufficient number of distance restraints to
impact prediction accuracy. The discriminative power, quantified through the enrichment metric, for
the optimal cross-linker length was \num{2.1}, whereas it was \numlist{1.4;1.5;1.9;1.7} for the
shorter and longer cross-linkers, respectively (\Fref{fig:xlink_results}). For the proteins 1X91
and 3M1X, the optimal cross-linker length did not yield any cross-links with a sequence separation
of at least ten and therefore did not provide relevant structural information. In those cases
protein structure prediction with longer cross-linker lengths provided better results. By combining
restraints obtained for all five cross-linker lengths, the average enrichment value could be
improved to \num{2.4}.

\subsection{Combination of cross-linkers with different reactivities results in improvements larger
  than seen when varying the spacer lengths}

In order to obtain valuable restraints for \emph{de novo} protein structure prediction a maximum
number of \gls{sse} pairs needs to be cross-linked. The availability of Lys-Asp/Glu reactive
cross-linkers allows for a better sequence coverage and therefore a wider coverage of \gls{sse}
pairs. Cross-links with different spacer lengths were simulated for the proteins in the benchmark
set using Xwalk. To assess the influence of Lys-Asp/Glu reactive cross-linkers on protein structure
prediction the same protocol was applied as for the Lys-Lys reactive cross-linkers. For the Lys-Glu
reactive cross-linkers a prediction accuracy of \SI{5.7}{\angstrom} and enrichment of \num{2.2} on
average could be achieved, which is comparable to the results for the Lys-Lys reactive
cross-linkers.

While Lys-Asp reactive cross-linkers also achieve improvements in prediction accuracy and
enrichment when compared to protein structure prediction without restraints, the results are
slightly worse than for Lys-Lys reactive cross-linkers with an average prediction accuracy of
\SI{6.0}{\angstrom} versus \SI{5.6}{\angstrom} and an average enrichment of \num{1.7} versus
\num{2.1} (\Fref{tab:xlink_prediction_results}). To a large part, the difference in the overall
results is caused by the results for the proteins 1ES9, 1T3Y, and 3M1X for which Lys-Asp reactive
cross-linkers failed to yield restraints between \gls{sse} pairs and there- fore failed to reduce
the conformational space significantly. Besides deviations regarding the average improvements over
all proteins, the spacer length deemed optimal also provides the best results for Lys-Asp/Glu
reactive cross-linking. Combining the restraints yielded for the optimal spacer lengths with
Lys-Lys/Asp/Glu reactive cross-links improves the sampling average sampling accuracy to
\SI{5.1}{\angstrom} and the average enrichment to \num{2.6}. Combining the restraints yielded by
all spacer lengths and cross-linker reactivities failed to further improve prediction results.

\section{Discussion}

\subsection{Prediction of the optimal cross-linker spacer length}

It has been shown frequently that chemical cross-linking data can be used to guide \emph{de novo}
structure prediction and selection of native-like models. Surely, the sensitivity, the broad
applicability to almost all proteins, the nearly physiological experimental condition during the
chemical cross-linking reaction, and the potential of automation are the main advantages for using
\gls{xl}-\gls{ms} to generate such restraints. However, the small number and high uncertainty of
restraints from chemical cross-links limit impact on \emph{de novo} proteins structure prediction,
in particular when compared to more data rich methods such as \gls{nmr} spectroscopy
\autocite{Weiner2014}.

One major limitation is the fact that distances between the functional groups in long and flexible
amino acid side-chains are measured. Therefore, a significant uncertainty is added to the
cross-linker length when converting \gls{xl}-\gls{ms} data into
$\mathrm{C_\beta}$-$\mathrm{C_\beta}$ restraints. A second challenge of chemical cross-links is
that only the maximum distance is restricted, but no information is obtained on the minimum
distance or the favored distance distribution. In result, even a ``zero length'' cross-linker
restricts the $\mathrm{C_\beta}$-$\mathrm{C_\beta}$ distance to the sum of the lengths of the two
connected side-chains (e.g. for Lys-Lys cross-links \SI{9.1}{\angstrom}).

In most of the cross-linking experiments, lysine residues are targeted. Lysines are excellent
targets because of their overrepresentation on protein surfaces and the clean chemistry of amine
modification. Nevertheless, their frequency is on average only about \SI{7}{\percent}. As a
consequence the number of cross-links, which can be formed for example in a \SI{25}{\kilo \dalton}
protein with a standard homobifunctional Lys-Lys-reactive cross-linking reagents with a spacer
length of \SI{6.4}{\angstrom} (length of DST) are in the range of about \num{20}. Only a small
fraction of these restraints will substantially limit the conformational space for the
protein. This number is usually too small to restrict the conformational space to an unambiguous
single protein fold. To increase the number of restrains it is possible to use cross-linkers with
longer spacer length or target amino acids such as Asp, Glu, Tyr, Ser, Thr, Arg, or Cys.

Restraints obtained with longer cross-linking reagents are less restrictive to the conformational
space. To evaluate the value of cross-links for protein structure prediction we determined for each
sequence distance (\numrange{0}{60} amino acids) how long a cross-linker has to be to link the
target amino acids. For example two lysines, which are separated by eight amino acids in sequence
were found to be linkable in all \num{3488} cases by a homobifunctional cross-linker with a length
of \textgreater \SI{30}{\angstrom} (as it is in BS(PEG)9). In our study, we stated the hypothesis
that it would be desirable if two target amino acids can only be linked in \SI{50}{\percent} of all
models created meaning that \SI{50}{\percent} of all structures could be discarded based on a
single cross-link. For example, for two lysines separated by \num{10} amino acids this would be the
case for cross-linker lengths of \SI{14.8}{\angstrom} (distance distributions for other amino acids
distances are shown in \fref{fig:xlink_res_pair_distributions}). Cross-links, which could only be
formed in less than \SI{50}{\percent} for the corresponding sequence distance, were considered as
being valuable. Based on this definition for all \num{2055} structures of the applied non-redundant
protein structure database the optimal spacer length was calculated. With this optimal spacer
length, the number of structural valuable cross-links has been maximized taking into account that
in general for modeling approaches few distance restraints of highly discriminative character are
less favorable than a higher number with a smaller discriminative power \autocite{Leitner2010,
  Havel1979}.

Since the optimal cross-linker length should depend on the protein size in a cubic root fashion to
convert volume into distance, it is not unexpected that this was also observed for the dependency
on the \gls{mw} (\Fref{fig:xlink_reactivity_lengths}). However, one has to keep in mind that the
formula might not be applicable to non-globular proteins and multi-domain proteins. However, in
case of multidomain proteins this formula should be applicable to the separated
domains. Remarkably, based on our simulation for proteins with \glspl{mw} of
\SIlist{10;25;50;100}{\kilo \dalton} the recommended spacer lengths are
\SIlist{9.0;11.5;13.9;17.0}{\angstrom}, respectively, which is quite close to the homobifunctional
amine-reactive commercially available cross-linkers DSG (\SI{7.7}{\angstrom}), BS3
(\SI{11.4}{\angstrom}), and EGS (\SI{16.1}{\angstrom}), which are currently the preferred choice to
study small (\textless \SI{20}{\kilo \dalton}), medium (\SIrange{20}{50}{\kilo \dalton}), and large
proteins (\textgreater \SI{50}{\kilo \dalton}), respectively.

\begin{figure}[h]
  \centering
  \includegraphics{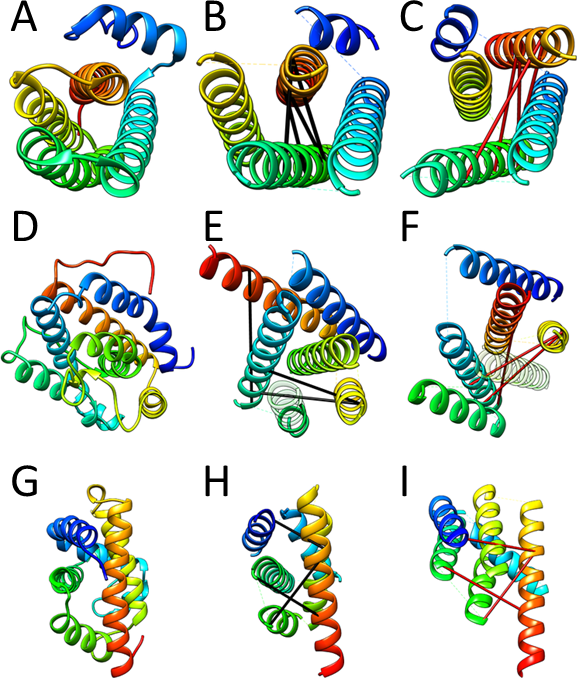}
  \caption[Selected prediction results from cross-linking data]{\textbf{Selected prediction results
      from cross-linking data.} Most accurate models sampled with and without using cross-linking
    restraints. The \gls{rmsd100} values of the most accurate models sampled for 1X91, 1J77, and
    1MBO were \SIlist{4.8;6.8;7.1}{\angstrom}. By using restraints yielded by Lys-Lys/Asp/Glu
    reactive cross-linkers, the accuracy could be improved to
    \SIlist{2.7;5.0;4.2}{\angstrom}. Shown are the native structures of 1X91, 1J77, and 1MBO (A, D,
    G), the most accurate models sampled without cross-linking restraints (B, E, H), and the most
    accurate models sampled with cross-linking restraints (C, F, I). Selected restraints are shown
    that are not fulfilled in the model predicted without cross-linking data (red bars), but that
    are fulfilled in the model predicted with cross-linking data (black bars).}
  \label{fig:xlink_gallery}
\end{figure}

Addressing different functional groups is a second approach to increase the total number of
distance restraints. The consequence is that the cross-linking reaction is either less effective or
specific (Asp, Glu, Tyr, Ser, Thr) creating challenges in data interpretation or the target amino
acids are less frequent (Arg and Cys) limiting the number of restraints observed. However, using
the same theoretical approach revealed that optimal spacer length for heterobifunctional Lys-Asp
and Lys-Glu cross-linker (\Fref{fig:xlink_gallery}) as well as homobifunctional Cys-Cys and Arg-Arg
cross-linker can be predicted with the same equation:
$l_{\mathit{opt}} [\si{\angstrom}] = k \times \sqrt[3]{\mathit{MW}} + \sqrt[3]{\mathit{SS1} +
  \mathit{SS2}}$ with $k \approx \frac{1}{3}$ in which $\mathit{SS1}$ and $\mathit{SS2}$ are the
average lengths of the cross-linked side-chains.

Comparing the two approaches to increase the number of valuable cross-links, it should be pointed
out that using several cross-linking reagents with different reactivities results in significantly
higher improvement of the model quality than using only lysine reactive cross-linking reagent but
with different spacer length.

\subsection{Challenges in using cross-linking data to guide \emph{de novo} modeling}

To test whether the cross-linker with the predicted optimal spacer length indeed perform best in
modeling we have chosen a \emph{de novo} structure prediction approach BCL::Fold for testing. Even
though comparative modeling using known protein structures as a template usually performs better
then \emph{de novo} modeling, our goal was to maximize impact of \gls{xl}-\gls{ms} restraints.

A major limiting factor for \emph{de novo} protein structure prediction methods is the vast size of
the conformational space. Cross-linking restraints can aid the computational prediction of a
protein’s tertiary structure by drastically reducing the size of the sampling space. Cross-linking
experiments yield a maximum Euclidean distance between the cross-linked residues, which increases
the sampling density in the relevant part of the conformational space.

A major limitation of using cross-linking restraints to guide protein structure prediction when
compared to restraints obtained from \gls{epr} and \gls{nmr} spectroscopy is that the cross-linker
length cannot be directly translated into a Euclidean distance between the cross-linked
residues. While cross-link prediction and evaluation methods like Xwalk \autocite{Kalkhof2005} are
successful at predicting if a certain cross-link can be formed in a given structure, explicit
modeling approaches are computationally too expensive for usage in a rapid scoring function
required for protein structure prediction. Approximations, such as the great circle on a sphere
presented here, are fast to compute but associated with increased uncertainties. Most of the
cross-linkers used can cover a long Euclidean distance and therefore the yielded restraints can be
fulfilled by a wide variety of conformations. In spite of this, cross-linking restraints display
some potential in limiting the size of the sampling space, resulting in a higher density of
accurate models. Further, the geometrical restraints derived from \gls{xl}-\gls{ms} allow for the
discrimination of a significant fraction of models representing incorrect topologies and therefore
improve the discriminative power of the scoring function.

\subsection{Abilities and limitations of protein structure prediction from limited experimental
  data}

We showed that incorporation of cross-linking data into a \emph{de novo} protein structure
prediction method improves the accuracy of the structure prediction. The two major challenges of
\emph{de novo} predictions are the sampling of structures as well as the discrimination of
inaccurate structures. In this study reduction of the conformational space was achieved through the
assembly of predicted \glspl{sse} with limited flexibility and the incorporation of geometrical
restraints derived from cross-linking data. The discrimination of inaccurate models is performed
through a scoring function which approximates the free energy. Assuming that the native structure
is in the global energy minimum, complete sampling and an accurate methods to measure free energy
would lead to the correct identification of the native conformation. However, the conformational
space is too large to be extensively sampled and the free energy needs to be approximated, which
results in ambiguity regarding the model which is most similar to the native
structure. Incorporating cross-linking data provides geometrical restraints, which can be used as
additional criteria to discriminate inaccurate models. While an average sampling accuracy of
\SI{5.1}{\angstrom}, when using restraints yielded by \gls{xl}-\gls{ms}, is a significant
improvement over the \SI{6.6}{\angstrom}, when not using cross-linking data at all, only for four
proteins it was possible to sample models with an \gls{rmsd100} of less than \SI{4}{\angstrom} when
compared to the crystal structure. Cross-linking data yields an upper boundary for the Euclidean
distance of the cross-linked residues, which allows for the placement of the second residue within
a sphere of volume $\frac{4}{3} \pi r^3$ around the first residue. Depending on the cross-link
distribution, topologically different models can fulfill the same restraint set. Discrimination
among those models is impossible with \gls{xl}-\gls{ms} restraints.

\subsection{Comparison of experimental and \emph{in silico} cross-links}

In order to draw general conclusion based on the analysis of hundreds of different structures this
study relies mainly on virtual cross-linking experiments. Unfortunately, although extensive
\gls{xl}-\gls{ms} data sets have been published for several proteins, it proved difficult to obtain
suitable experimental data sets for the present benchmark due to additional requirements:
\begin{inparaenum}[(i)]
  \item the protein must be monomeric and small enough for \emph{de novo} protein folding with
    BCL::Fold,
  \item an experimental atomic detail structure for comparison, and
  \item a large data set of intramolecular cross-links must be available.
\end{inparaenum}
Results for the four cases P11, FGF2, cytochrome c, and oxymyoglobin that came closest are reported
to demonstrate our efforts to work not only with simulated data. However, for P11 and FGF2 using
experimentally determined restraints did not improve the prediction results in a statistically
significant way. For P11, only three restraints were available with a maximum sequence separation
of nine residues. Because of the small sequence separation, these restraints contain very limited
structural information and no improvement in \emph{de novo} folding can be expected. The tertiary
structure of FGF2 contains twelve $\mathrm{\beta}$-strands with several $\mathrm{\beta}$-strands
that are strongly bent. This protein is too large for \emph{de novo} structure determination with
BCL::Fold. As BCL::Fold is unable to sample the conformation of the protein in the first place, no
significant improvement was expected or observed when \gls{xl}-\gls{ms} data were
added. Nevertheless, the value of the predicted cross-links in comparison to experimental
cross-links could be validated with the two proteins cytochrome c and oxymyoglobin for which
experimental cross-links had been published in the \gls{xl} database
\autocite{Petrotchenko2010}. For cytochrome c (\gls{pdb} entry 1HRC), we indeed found that the
cross-linker with predicted optimal spacer length of \SI{10.2}{\angstrom} performed best. However,
for oxymyoglobin (\gls{pdb} entry 1MBO) the longer spacers improved the accuracy slightly more than
the cross-linker with the optimal spacer length. Interestingly, on the one hand for both proteins
several cross-links, which should be possible, could not be detected, which might be due to
experimental or analytical reasons. On the other hand, also several cross-links, which were
experimentally, identified which were not predicted. An examination of these data revealed that
most of these cross-links are not present in the virtual data set because their
$\mathrm{C_\beta}$-$\mathrm{C_\beta}$ distances exceed the expected maximum length. This finding is
in agreement with Merkley \emph{et al.}~\autocite{Merkley2014}, who evaluated protein structures by
molecular dynamics and reported that usually a high number of experimental approved cross-links
exceed the theoretical maximal spatial distance due to structure flexibility. It was concluded for
the investigation of Lys-Lys distances using a BS3/DSS cross-linking reagent an upper bound of
\SIrange{26}{30}{\angstrom} for $\mathrm{C_\alpha}$-atoms \autocite{Merkley2014}.

On the other hand, spacer conformations usually adapt lengths that are somehow distributed between
their minimal and maximal lengths. In line it was also reported that many spacers in commercially
available cross-link agents preferable adopt conformations, which are significantly below the cited
maximal spacer length \autocite{Green2001}. Thus, ideally cross-linking results should be evaluated
based on experimentally derived or simulated ensembles of in-solution structures instead of using
X-ray structures as reference. However, to address all degrees of flexibility during the \emph{de
  novo} structure prediction is currently too resource intensive. Furthermore, there are many
additional practical challenges, which may prevent the formation or identification of cross-links,
and thus may result in more meaningful results using a cross-linker with a non-optimal
length. Nevertheless, for both structures the sampling accuracies could also be improved by
\SI{0.7}{\angstrom} based on the experimentally determined restraints, which is only slightly worse
than the improvement of \SI{1.0}{\angstrom} observed based on \emph{in silico} cross-links.

\section{Conclusion}

Recent development of high-resolution \gls{ms} instruments enables the analysis of proteins not
accessible to \gls{nmr} spectroscopy and X-ray crystallography. Data obtained from those
experiments can be translated into structural restraints to guide protein structure prediction. The
information content of a geometrical restraint obtained from \gls{xl}-\gls{ms} experiments is
directly dependent on the used spacer length. Thus, the choice of the spacer length is an important
step.

Firstly, for amino acids pairs close in sequence only minimum structural information is obtained if
the spacer is too long. Here we determine the optimal spacer length to gain structural information
on lysines with a sequence separation of $S$, we estimated a length as
$E = 5.5 \times ln(S) + 2.2$. Secondly, we demonstrate that for \emph{de novo} protein structure
prediction the optimal spacer length depends on the \gls{mw} of the protein of interest and the
length of the cross-linked side-chains ($\mathit{SS1}$ and $\mathit{SS2}$) and can be predicted as
$l_{\mathit{opt}} = k \times \sqrt[3]{\mathit{MW}} + \sqrt[3]{\mathit{SS1} + \mathit{SS2}}$, with
$k \approx \frac{1}{3}$.

We also demonstrate that restraints obtained from cross-linking experiments contribute moderately
to solving the major challenges of \emph{de novo} protein structure prediction --- the vast size of
the conformational space and discrimination of inaccurate models. Using restraints from
cross-linking experiments significantly increases the sampling density of native-like models and
contribute to the discrimination of incorrect models. By combining cross-linking restraints with
knowledge-based scoring functions, the average accuracy of the sampled models could be improved by
up to \SI{2.2}{\angstrom} and the average enrichment of accurate models could be improved from
\SI{11}{\percent} to \SI{24}{\percent}.

Conclusively, we believe this study can help in the planing of \gls{xl}-\gls{ms} experiments as
well as to evaluate how much information can be gained by \gls{xl}-\gls{ms} experiments and the
ambiguity that remains.

\section{Acknowledgments}

This study was supported by grants from Deutsche Forschungsgemeinschaft Transregio 67 (subproject
Z4) and ESF Investigator group GPCR 2. Work in the Meiler laboratory is supported through NIH (R01
GM080403, R01 GM099842, R01 DK097376) and NSF (CHE 1305874). This research used resources of the
Oak Ridge Leadership Computing Facility at the Oak Ridge National Laboratory, which is supported by
the Office of Science of the U.S. Department of Energy under Contract No. DE-AC05-00OR22725.

Parts of the data analysis were performed using the R package. The renderings of the models were
created using Chimera \autocite{Pettersen2004}.


\printbibliography

\section{Supplementary data}

\begin{table}
  \small
  \centering
  \begin{tabular}{lrrrrrrrrrr}
\toprule
  & \multicolumn{2}{c}{$\mathit{optimal}$} & \multicolumn{2}{c}{$\mathit{short1}$} & \multicolumn{2}{c}{$\mathit{short2}$} & \multicolumn{2}{c}{$\mathit{long1}$} & \multicolumn{2}{c}{$\mathit{long2}$} \\
\cmidrule{2-11}
Protein & length & \#rest & length & \#rest & length & \#rest & length & \#rest & length & \#rest \\
\midrule
1HRC    & \SI{10.2}{\angstrom}   & 13     & \SI{2.5}{\angstrom}    & 0      & \SI{7.5}{\angstrom}    & 7      & \SI{17.5}{\angstrom}   & 27     & \SI{30}{\angstrom}     & 107    \\
3IV4    & \SI{10.4}{\angstrom}   & 5      & \SI{2.5}{\angstrom}    & 2      & \SI{7.5}{\angstrom}    & 2      & \SI{17.5}{\angstrom}   & 7      & \SI{30}{\angstrom}     & 13     \\
1BGF    & \SI{10.7}{\angstrom}   & 6      & \SI{2.5}{\angstrom}    & 3      & \SI{7.5}{\angstrom}    & 4      & \SI{17.5}{\angstrom}   & 10     & \SI{30}{\angstrom}     & 13     \\
1T3Y    & \SI{10.9}{\angstrom}   & 35     & \SI{2.5}{\angstrom}    & 9      & \SI{7.5}{\angstrom}    & 20     & \SI{17.5}{\angstrom}   & 42     & \SI{30}{\angstrom}     & 63     \\
3M1X    & \SI{10.9}{\angstrom}   & 1      & \SI{2.5}{\angstrom}    & 0      & \SI{7.5}{\angstrom}    & 0      & \SI{17.5}{\angstrom}   & 5      & \SI{30}{\angstrom}     & 19     \\
1X91    & \SI{11.0}{\angstrom}   & 2      & \SI{2.5}{\angstrom}    & 0      & \SI{7.5}{\angstrom}    & 1      & \SI{17.5}{\angstrom}   & 8      & \SI{30}{\angstrom}     & 27     \\
1JL1    & \SI{11.2}{\angstrom}   & 7      & \SI{2.5}{\angstrom}    & 0      & \SI{7.5}{\angstrom}    & 3      & \SI{17.5}{\angstrom}   & 11     & \SI{30}{\angstrom}     & 24     \\
1MBO    & \SI{11.3}{\angstrom}   & 9      & \SI{2.5}{\angstrom}    & 0      & \SI{7.5}{\angstrom}    & 3      & \SI{17.5}{\angstrom}   & 23     & \SI{30}{\angstrom}     & 77     \\
2QNL    & \SI{11.5}{\angstrom}   & 6      & \SI{2.5}{\angstrom}    & 4      & \SI{7.5}{\angstrom}    & 4      & \SI{17.5}{\angstrom}   & 8      & \SI{30}{\angstrom}     & 15     \\
2AP3    & \SI{12.1}{\angstrom}   & 53     & \SI{2.5}{\angstrom}    & 0      & \SI{7.5}{\angstrom}    & 19     & \SI{17.5}{\angstrom}   & 136    & \SI{30}{\angstrom}     & 427    \\
1J77    & \SI{12.2}{\angstrom}   & 29     & \SI{2.5}{\angstrom}    & 7      & \SI{7.5}{\angstrom}    & 16     & \SI{17.5}{\angstrom}   & 36     & \SI{30}{\angstrom}     & 70     \\
1ES9    & \SI{12.5}{\angstrom}   & 8      & \SI{7.5}{\angstrom}    & 0      & \SI{17.5}{\angstrom}   & 1      & \SI{37.5}{\angstrom}   & 17     & \SI{45}{\angstrom}     & 20     \\
3B5O    & \SI{12.7}{\angstrom}   & 15     & \SI{7.5}{\angstrom}    & 2      & \SI{17.5}{\angstrom}   & 8      & \SI{37.5}{\angstrom}   & 21     & \SI{45}{\angstrom}     & 25     \\
1XQ0    & \SI{13.3}{\angstrom}   & 9      & \SI{7.5}{\angstrom}    & 0      & \SI{17.5}{\angstrom}   & 4      & \SI{37.5}{\angstrom}   & 14     & \SI{45}{\angstrom}     & 44     \\
2IXM    & \SI{13.5}{\angstrom}   & 41     & \SI{7.5}{\angstrom}    & 20     & \SI{17.5}{\angstrom}   & 41     & \SI{37.5}{\angstrom}   & 49     & \SI{45}{\angstrom}     & 57     \\
\bottomrule
\end{tabular}

  \caption[Lys-Lys cross-links yielded by different spacer
  lengths]{\textbf{Lys-Lys cross-links yielded by different spacer lengths.} Cross-links obtained
    for the benchmark proteins. Simulated and experimentally determined cross-links were obtained
    for the fifteen benchmark proteins. For each protein, an optimal spacer length was determined
    ($\text{optimal}$). Additional cross-links were simulated for two shorter ($\text{short1}$ and
    $\text{short2}$) and two longer ($\text{long1}$ and $\text{long2}$) spacer lengths. The number
    of yielded cross-links ($\#\text{rest}$) is shown for each spacer length. For the two proteins
    1HRC and 1MBO, experimentally determined cross-links were published.}
  \label{tab:xlink_num_lys_lys}
\end{table}

\begin{table}
  \small
  \centering
  \begin{tabular}{lrrrrrrrrrr}
\toprule
  & \multicolumn{4}{c}{Without restraints} & \multicolumn{2}{c}{Optimal Lys/Lys} & \multicolumn{2}{c}{All Lys/Lys} & \multicolumn{2}{c}{All reactivities} \\
\cmidrule{2-11}
Protein & $\mathit{best}$ & $\sigma_{\text{best}}$ & $e$   & $\sigma_e$  & $\mathit{best}$ & $e$   & $\mathit{best}$ & $e$   & $\mathit{best}$ & $e$   \\
\midrule
1HRC    & \SI{4.5}{\angstrom}  & \SI{0.3}{\angstrom}  & 0.8 & 0.1 & \SI{3.8}{\angstrom} & 2.0 & \SI{3.8}{\angstrom}  & 2.0 &  \SI{3.7}{\angstrom} & 5.9 \\
3IV4    & \SI{6.7}{\angstrom}  & \SI{0.2}{\angstrom}  & 1.2 & 0.3 & \SI{5.7}{\angstrom} & 2.5 & \SI{5.3}{\angstrom}  & 2.5 &  \SI{5.2}{\angstrom} & 1.9 \\
1BGF    & \SI{6.6}{\angstrom}  & \SI{0.4}{\angstrom}  & 1.0 & 0.2 & \SI{5.7}{\angstrom} & 2.1 & \SI{4.9}{\angstrom}  & 2.4 &  \SI{6.2}{\angstrom} & 1.6 \\
1T3Y    & \SI{7.0}{\angstrom}  & \SI{0.7}{\angstrom}  & 1.7 & 0.4 & \SI{6.4}{\angstrom} & 2.9 & \SI{5.7}{\angstrom}  & 3.0 &  \SI{6.2}{\angstrom} & 2.3 \\
3M1X    & \SI{3.8}{\angstrom}  & \SI{0.1}{\angstrom}  & 0.7 & 0.2 & \SI{3.8}{\angstrom} & 0.7 & \SI{3.6}{\angstrom}  & 1.5 &  \SI{3.6}{\angstrom} & 1.7 \\
1X91    & \SI{4.8}{\angstrom}  & \SI{0.2}{\angstrom}  & 2.0 & 0.5 & \SI{4.8}{\angstrom} & 2.0 & \SI{2.0}{\angstrom}  & 3.2 &  \SI{2.1}{\angstrom} & 3.5 \\
1JL1    & \SI{6.4}{\angstrom}  & \SI{0.4}{\angstrom}  & 1.2 & 0.1 & \SI{5.6}{\angstrom} & 2.1 & \SI{5.3}{\angstrom}  & 2.8 &  \SI{5.1}{\angstrom} & 2.7 \\
1MBO    & \SI{7.1}{\angstrom}  & \SI{0.6}{\angstrom}  & 0.8 & 0.3 & \SI{6.4}{\angstrom} & 2.0 & \SI{6.5}{\angstrom}  & 1.6 &  \SI{4.2}{\angstrom} & 2.5 \\
2QNL    & \SI{7.0}{\angstrom}  & \SI{0.6}{\angstrom}  & 1.0 & 0.3 & \SI{4.8}{\angstrom} & 1.9 & \SI{4.1}{\angstrom}  & 2.1 &  \SI{6.1}{\angstrom} & 2.1 \\
2AP3    & \SI{2.5}{\angstrom}  & \SI{0.1}{\angstrom}  & 1.6 & 0.5 & \SI{2.0}{\angstrom} & 3.0 & \SI{1.6}{\angstrom}  & 3.1 &  \SI{2.2}{\angstrom} & 2.0 \\
1J77    & \SI{6.8}{\angstrom}  & \SI{0.3}{\angstrom}  & 0.5 & 0.2 & \SI{5.0}{\angstrom} & 2.0 & \SI{4.0}{\angstrom}  & 2.4 &  \SI{3.8}{\angstrom} & 3.2 \\
1ES9    & \SI{7.3}{\angstrom}  & \SI{0.8}{\angstrom}  & 1.1 & 0.6 & \SI{5.7}{\angstrom} & 2.1 & \SI{5.6}{\angstrom}  & 2.8 &  \SI{6.3}{\angstrom} & 2.9 \\
3B5O    & \SI{9.2}{\angstrom}  & \SI{0.9}{\angstrom}  & 1.4 & 0.2 & \SI{8.6}{\angstrom} & 1.9 & \SI{9.0}{\angstrom}  & 2.6 &  \SI{7.1}{\angstrom} & 1.9 \\
1XQ0    & \SI{9.9}{\angstrom}  & \SI{1.0}{\angstrom}  & 1.1 & 0.3 & \SI{8.3}{\angstrom} & 1.9 & \SI{8.5}{\angstrom}  & 2.4 &  \SI{7.4}{\angstrom} & 2.1 \\
2IXM    & \SI{9.4}{\angstrom}  & \SI{0.9}{\angstrom}  & 1.1 & 0.4 & \SI{7.9}{\angstrom} & 1.7 & \SI{8.5}{\angstrom}  & 1.7 &  \SI{7.0}{\angstrom} & 1.9 \\
\midrule
Ø       & \SI{6.6}{\angstrom}  & \SI{0.5}{\angstrom}  & 1.1 & 0.3 & \SI{5.6}{\angstrom} & 2.1 & \SI{5.2}{\angstrom}  & 2.4 &  \SI{5.1}{\angstrom} & 2.6 \\
\bottomrule
\end{tabular}

  \caption[Protein structure prediction results for different cross-linker
  reactivities]{\textbf{Protein structure prediction results for different cross-linker
      reactivities.} Comparison between structure prediction results with and without cross-linking
    restraints. By using geometrical restraints obtained from cross-linking experiments, the size
    of the sampling space can be reduced resulting in an improved sampling accuracy. This is shown
    by significant improvements in the \gls{rmsd100} value of the most accurate model
    ($\text{best}$). Furthermore, cross-linking restraints provide geometrical information, which
    improves the discrimination power of the scoring function, leading to an improvement in the
    enrichment (e). Without restraints, ten independent prediction trajectories were conducted and
    the standard deviation of the accuracy of the best model ($\sigma_{\mathit{best}}$) and the
    enrichment ($\sigma_e$) are reported.}
  \label{tab:xlink_prediction_results}
\end{table}

\end{document}